\documentclass[useAMS,usenatbib]{mn2e}

\renewcommand\sfdefault{ptm}
\usepackage{sfmath}

\usepackage{lscape}
\usepackage{aas_macros}

\usepackage{epsfig}
\usepackage{natbib}

\newbox\grsign \setbox\grsign=\hbox{$>$} \newdimen\grdimen \grdimen=\ht\grsign
\newbox\simlessbox \newbox\simgreatbox \newbox\simpropbox \newbox\wtildebox 
\setbox\simgreatbox=\hbox{\raise.5ex\hbox{$>$}\llap
    {\lower.5ex\hbox{$\sim$}}}\ht1=\grdimen\dp1=0pt
\setbox\simlessbox=\hbox{\raise.5ex\hbox{$<$}\llap
    {\lower.5ex\hbox{$\sim$}}}\ht2=\grdimen\dp2=0pt

\newcommand{\be}{\mbox{\begin{equation}}}
\newcommand{\ee}{\mbox{\end{equation}}}

\newcommand{\Cref}{\mbox{$m_{\rm ref}$}}

\newcommand{\msun}{\mbox{${\rm M}_\odot$}}

\renewcommand{\d}{{\rm d}} 

\title{The initial mass spectrum of old globular clusters in dwarf galaxies}   
\author{J.~M.~Diederik Kruijssen\thanks{kruijssen@mpa-garching.mpg.de} and Andrew~P.~Cooper\thanks{acooper@mpa-garching.mpg.de}\\
Max-Planck Institut f\"{u}r Astrophysik, Karl-Schwarzschild-Stra\ss e 1, 85748, Garching, Germany}
\begin{document}

\date{Accepted 2011 October 19. Received 2011 October 18; in original form 2011 September 9.}

\pagerange{\pageref{firstpage}--\pageref{lastpage}} \pubyear{2011}
\label{firstpage}

\maketitle

\begin{abstract}
{We test whether the masses of old globular clusters (GCs) in dwarf galaxies are consistent with the same initial mass spectrum as young massive clusters (YMCs) in nearby star-forming galaxies. The most massive GCs of dwarf galaxies are compared to their expected masses when drawing from the Schechter-type ICMF of YMCs. It is found that the most massive GCs of galaxies in the stellar mass range $M_{\star,{\rm gal}}=10^7$--$10^9~\msun$} are consistent with the same initial mass spectrum as YMCs in about 90\% of the cases, suggesting that their formation mechanisms were the same. For the remaining 10\%, the most massive clusters are nuclear GCs, which have been able to grow to higher masses through further merging after their initial formation ended. Because the effects of cluster disruption are weaker for more massive clusters, we estimate that up to one third of the metal-poor GCs in the Milky Way may have a nuclear origin, while the remaining two thirds formed through the same process as YMCs in the local universe. A log-normal ICMF is inconsistent with observed GCs at a 99.6\% confidence level.
\end{abstract}

\begin{keywords}
galaxies: evolution -- galaxies: dwarf galaxies -- galaxies: starburst -- galaxies: star clusters -- globular clusters: general
\end{keywords}

\renewcommand{\thefootnote}{\mbox{$^\arabic{footnote}$}}
\setcounter{footnote}{0}

\section{Introduction} \label{sec:intro}
The quest to understand how globular clusters (GCs) formed goes back several decades \citep[e.g.][]{searle78,fall85}. Recent efforts have concentrated on the formation sites and mechanisms of GCs, driven by the question whether GCs are presently still forming in the nearby universe. The Galactic GC population likely has its roots in several environments \citep{mackey04,muratov10}. Metal-poor GCs are thought to originate from dwarf galaxies that were accreted by the Milky Way \citep{prieto08,penarrubia09,lee09}, while metal-rich GCs may have formed in gas-rich galaxy mergers \citep{ashman92,whitmore99,zepf99} or unstable galaxy discs \citep{shapiro10}. Whichever environment GCs have formed in, their masses suggest it should have been in a burst of star formation, potentially similar to nearby galaxy mergers \citep{holtzman92,schweizer96} or starburst dwarf galaxies \citep{adamo10}.

A starburst origin for GCs is required if their initial cluster mass function (ICMF) was the same as that of young star cluster populations in the local universe, which follows a power law or \citet{schechter76} type distribution with index $-2$ down to some lower mass limit \citep{portegieszwart10}. The random sampling from such an ICMF requires a high star formation rate (SFR) for the production of massive clusters, unless the lower mass limit exceeds $\sim10^5~\msun$, which is not consistent with young cluster populations. Indeed, in nearby spiral galaxies, starbursts and galaxy mergers, the masses of young massive clusters (YMCs) are all consistent with an approximately universal ICMF \citep{gieles06b,larsen09,portegieszwart10}, except for an environmentally dependent variation of its upper truncation \citep{bastian11b}. The approximately log-normal present-day mass function of GCs also bears some traits of formation according to a Schechter-type ICMF \citep{harris94,mclaughlin96,burkert00,jordan07}, but this necessarily implies that the vast majority of low-mass GCs has been destroyed due to tidal disruption \citep{fall77,elmegreen97,vesperini01,fall01,kruijssen09b}. This scenario contrasts with the idea that the current mass function of GCs would reflect a fundamentally different (e.g. log-normal) ICMF \citep{vesperini03,parmentier07}{, which has been put forward as an alternative explanation for the approximate universality of the peaked present-day GC mass function. A similar ICMF for GCs and young clusters in nearby galaxies would imply that the disruption histories of GCs have also been close to universal. Considering the broad range of galactic environments in which GCs are currently found, this suggests that most of the GC disruption may have occurred at the epoch of their formation (\citealt{elmegreen10,kruijssen11}, Kruijssen et al. in prep.), when their environments were likely more comparable.}

It is relevant to ask whether or not the ICMFs of GCs and young clusters in the local universe are the same, because the ICMF is a tracer of the cluster formation process \citep{elmegreen96}. A similar ICMF would thus be required if GCs and YMCs are two stages of the same type of object in terms of their physics and formation mechanism. Such evidence cannot be found for the entire GC mass range, due to the ongoing disruption of (low-mass) stellar clusters, but this problem can be addressed using the most massive GCs. Given an ICMF, the mass of the most massive star cluster can be predicted statistically if a fraction of star formation resulting in bound clusters is assumed, i.e. the cluster formation efficiency or CFE. This has been done for YMCs \citep{bastian08}, but for GCs such an approach has been obstructed by unknowns about the conversion of current GC properties to their original form. 

We aim to connect YMCs and GCs by adopting the recent advancements made in the understanding of YMC formation in the local universe. These will be used to predict the expected masses of the most massive GCs in dwarf galaxies for limiting cases, to verify whether they are consistent with random sampling from the same ICMF as YMCs. This should provide an indication of whether or not their formation mechanisms are compatible. Dwarf galaxies are the most suitable for such an analysis, because they have not accreted as much since GC formation as massive galaxies have. Moreover, the masses of the most massive YMCs \citep[e.g.][]{bastian06} exceed the stellar masses of the faintest dwarf galaxies \citep[e.g.][]{mateo98}, and therefore dwarf galaxies are the prime targets in which the limits of GC formation can be explored. This naturally constrains our analysis to the origin of metal-poor GCs.

\section{The relation between globular cluster mass and galaxy stellar mass} \label{sec:method}
Young star clusters in the Milky Way and nearby galaxies follow a power law ICMF with an exponential truncation \citep[cf.][]{schechter76}, over a mass range of $M=10^2$--$10^8~\msun$ \citep{maraston04,gieles06b,portegieszwart10}. The probability density of cluster masses $\xi(M)$ is expressed as
\begin{equation}
\label{eq:ICMF}
\xi(M)\d M\propto M^{-2}\exp{(-M/M_{\rm c})}\d M ,
\end{equation}
where $M_{\rm c}$ represents the truncation mass and $\xi(M)$ is normalised such that $\int\xi(M){\rm d}M=1$, integrated between the physical mass limits $M_{\rm MIN}$ and $M_{\rm MAX}$. In this study, we adopt $M_{\rm MIN}=10^2~\msun$. The physical upper mass limit is set equal to the stellar mass of the host galaxy, i.e. $M_{\rm MAX}=M_{\star,{\rm gal}}$, with generally $M_{\rm c}<M_{\rm MAX}$.

When randomly sampling cluster masses from an ICMF, the probability distribution function for the most massive sampled cluster $p(M_{\rm max})$ is given by \citep{maschberger08}:
\begin{equation}
\label{eq:pmmax}
p(M_{\rm max})=N\left(\int_{M_{\rm MIN}}^{M_{\rm max}}\xi(M'){\rm d}M'\right)^{N-1}\xi(M_{\rm max}) ,
\end{equation}
where $N$ is the total number of clusters. It is obtained by integration of Eq.~\ref{eq:ICMF} after renormalising the ICMF to match the total stellar mass formed in clusters $M_{\star,{\rm cl}}$. The expected $M_{\rm max}$ is given by the median of Eq~\ref{eq:pmmax}, with the 16th and 84th percentiles spanning one standard deviation up and down. We use Eq.~\ref{eq:pmmax} to compute the most massive cluster as a function of host galaxy stellar mass, and compare the result to observed GCs and (for verification) YMCs.

The comparison between the theoretically expected most massive GC and observed GCs requires certain assumptions about the following quantities and their dependences on the host galaxy.
\begin{itemize}
\item[(1)]
The fraction of the present-day galaxy stellar mass that was formed coevally with the GCs ($f_{\star,{\rm old}}$).
\item[(2)]
The fraction of star formation that yields bound stellar clusters, i.e. the CFE ($\Gamma$). 
\item[(3)]
The exponential truncation mass of the ICMF ($M_{\rm c}$).
\item[(4)]
The mass loss suffered by GCs since their formation, parametrized by the disruption time-scale of a $10^5~\msun$ cluster ($t_5$).
\end{itemize}
The parameters $f_{\star,{\rm old}}$ and $\Gamma$ relate the current stellar mass of the galaxy $M_{\star,{\rm gal}}$ to the normalisation of the ICMF through $M_{\star,{\rm cl}}= \Gamma  f_{\star,{\rm old}}M_{\star,{\rm gal}}$, and $t_5$ determines what the present-day mass of the GCs is, after a Hubble time of dynamical evolution (see below).

The hypothesis that old GCs and YMCs form from an ICMF with a common functional form can be tested to first order by setting the above quantities to values that enable the formation of the most massive clusters. Such a limiting case would assume that (1) all present stellar mass of the galaxy was formed at the same time as the GCs ($f_{\star,{\rm old}}=1$), (2) that all of its stars formed in clusters ($\Gamma=1$), (3) that the ICMF is a pure power law with an upper mass limit equal to the stellar mass of the galaxy ($M_{\rm c}=\infty$ and $M_{\rm MAX}=M_{\star,{\rm gal}}$), and (4) that there has been no mass loss by dynamical processes ($t_5=\infty$). Stellar evolution would still apply and is included as a fixed, 30\% decrease of the cluster mass over a Hubble time. The corresponding parameter set is shown as the `extreme mode' in Table~\ref{tab:parameters} and gives an upper limit to the relation between the most massive GC and the stellar mass of the host galaxy, provided that the GCs are randomly sampled from a \citet{schechter76} type ICMF. If any of the observed GCs lies above this relation, it would imply that at least some GCs are too massive to have been sampled from the ICMF of YMCs, and would suggest that they formed through a different mechanism.

\begin{figure}
\center\resizebox{8cm}{!}{\includegraphics{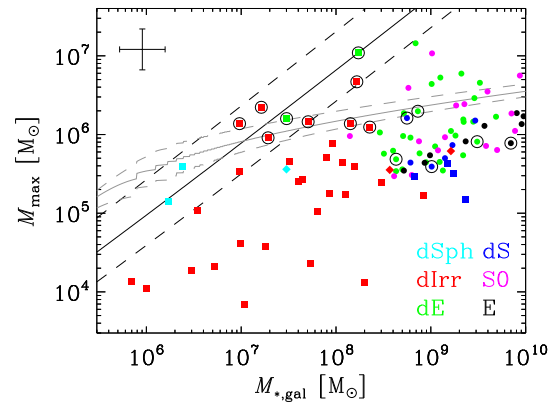}}\\
\caption[]{\label{fig:mgcmax0}\sf
      Mass of the most massive GC versus the stellar mass of its host galaxy. The black solid line shows the theoretical upper limit (see text) to the relation for a power law ICMF, with the standard deviation due to random sampling from Eq.~\ref{eq:pmmax} indicated as dashed lines. Grey lines denote the same for a log-normal ICMF (see text). Coloured symbols denote observed GCs, with host galaxy types indicated by the legend. Squares mark GCs from nearby field dwarf galaxies \citep{georgiev10}, dots denote GCs from galaxies in the Virgo Cluster \citep{cote04,peng08,jordan09}, and diamonds indicate (in order of increasing $M_{\star,{\rm gal}}$) Fornax \citep{mackey03c,lokas09}, the SMC \citep{mackey03b} and the LMC \citep{mackey03}. Black circles mark nuclear GCs, defined as having a projected galactocentric radius $<0.5$~kpc. The typical error margin on the data is shown in the top left corner.
                 }
\end{figure}
The `extreme mode' relation between the most massive GC and host galaxy stellar mass is shown in Fig.~\ref{fig:mgcmax0}, together with the observed most massive GCs from nearby field dwarf galaxies and galaxies in the Virgo Cluster. For reference, we have also included the relation for a { purely} log-normal ICMF, with a peak mass $\log{(M_{\rm peak}/\msun)}\sim5.1$, dispersion $\sigma\sim0.6$ \citep{vesperini03}, and a weak dependence of both on galaxy mass as in \citet{jordan07}. Figure~\ref{fig:mgcmax0} shows that even in this limiting case a log-normal ICMF is not consistent with the observed GC masses. { This is mainly caused by the steep decline of a log-normal function at the high-mass end.} By contrast, none of the observed GCs has a mass higher than the theoretical upper limit for randomly sampling the most massive GC from a Schechter-type ICMF. This shows that the initial, total star-forming potential of the galaxy could indeed `support' the formation of such massive GCs through the initial mass spectrum we see in the local universe. It does not yet imply that the ICMFs of GCs and YMCs are fully consistent, because this would assume that the adopted parameter set represents the conditions in real galaxies.

However, the scenario sketched as the `extreme mode' of GC formation in Fig.~\ref{fig:mgcmax0} is not quite realistic. Only a certain fraction of the stellar content of dwarf galaxies formed coevally with GCs \citep[e.g.][]{mcquinn10}, there is no evidence for such efficient cluster formation that $\Gamma=1$ \citep[e.g.][]{goddard10,bressert10,silvavilla11}, there is likely an environmentally dependent upper mass limit to cluster formation \citep[e.g.][]{gieles06b,bastian08}, and GCs are known to have lost at least some fraction of their initial mass due to dynamical evolution \citep{fall01}. The values of these parameters and their relation to the properties of the host galaxy have been constrained empirically for cluster formation in the local universe. We adopt these relations to test whether GC formation has proceeded in a way that is consistent with the ICMF of young clusters in nearby galaxies.

In a study of seven nearby star-forming galaxies, \citet{goddard10} related $\Gamma$ to the star formation rate density $\Sigma_{\rm SFR}$:
\begin{equation}
\label{eq:cfe}
\Gamma=0.29\left(\frac{\Sigma_{\rm SFR}}{\msun~{\rm yr}^{-1}~{\rm kpc}^{-2}}\right)^{0.24} ,
\end{equation}
which has been confirmed by \citet{adamo10,adamo11}. The relation holds over $\Sigma_{\rm SFR}=6\times10^{-3}$--1~$\msun~{\rm yr}^{-1}~{\rm kpc}^{-2}$, corresponding to $\Gamma=0.08$--0.29. \citet{bastian08} explored the relation between the brightest cluster in a galaxy and the star formation rate. If the CFE is only allowed to vary slowly as in Eq.~\ref{eq:cfe}, then $M_{\rm c}$ must increase with the SFR. Based on his results on YMCs, we assume
\begin{equation}
\label{eq:mstar}
M_{\rm c}= 7\times10^5\left(\frac{{\rm SFR}}{\msun~{\rm yr}^{-1}}\right)^{0.5}~\msun,
\end{equation}
which reproduces the typical $M_{\rm c}\sim 2\times10^5~\msun$ found by \citet{larsen09} for spiral galaxies with ${\rm SFR}\sim0.1~\msun~{\rm yr}^{-1}$ and $M_{\rm c}\sim 3\times10^6~\msun$ for the Antennae galaxies, assuming ${\rm SFR}=20~\msun~{\rm yr}^{-1}$ \citep{zhang01}. Equations~\ref{eq:cfe} and~\ref{eq:mstar} depend on the galaxy stellar mass $M_{\star,{\rm gal}}$ through SFR$(M_{\star,{\rm gal}})$ and $\Sigma_{\rm SFR}(M_{\star,{\rm gal}})$ (see below).

After their formation, star clusters lose mass due to stellar evolution and tidal disruption. While this may not be very important for YMCs, the mass of GCs will have decreased after almost a Hubble time of evolution. A simple empirical relation for the mass evolution of a cluster was derived by \citet{lamers05}:
\begin{equation}
\label{eq:lamers}
M(t)=M_{\rm i}\left[\mu_{\rm se}-\frac{\gamma t}{t_5}\left(\frac{M_{\rm i}}{10^5~\msun}\right)^{-\gamma}\right]^{1/\gamma} ,
\end{equation}
which depends on the initial cluster mass $M_{\rm i}$, the stellar evolutionary mass loss \citep[$\mu_{\rm se}\sim 0.7$, e.g.][]{lamers05}, the parameter $\gamma$, which sets the mass-dependence of the disruption time \citep[we adopt a value of $\gamma\sim0.7$, see][]{lamers10}, and the disruption time-scale of a $10^5~\msun$ cluster $t_5$, which is environmentally dependent \citep[e.g.][]{kruijssen11,bastian11}. In their analysis of GCs in the Virgo Cluster, \citet{jordan07} find that the peak of the GC luminosity function (GCLF) depends on the magnitude of the host galaxy. By assuming that the peak luminosity is approximately proportional to the disruption time-scale \citep{gieles09} and by using a constant $M/L$ ratio to convert the galaxy magnitude to its total stellar mass, this can be translated to a very weak dependence of $t_5$ on $M_{\star,{\rm gal}}$:
\begin{equation}
\label{eq:t5}
t_5=C_{\rm dis}\left(\frac{M_{\star,{\rm gal}}}{10^9~\msun}\right)^{-0.1} ,
\end{equation}
where $\log{(C_{\rm dis}/{\rm yr})}\sim10.5$ for SMC/LMC-type galaxies with $M_{\star,{\rm gal}}\sim 10^9~\msun$ \citep{lamers05a}. For the Milky Way, this gives $t_5\sim 2\times 10^{10}$~yr, consistent with the range derived for Galactic GCs \citep{kruijssen09}. The scatter among GCs in a particular galaxy is non-negligible though, and we therefore assume a spread of a factor of two, i.e. $\log{(C_{\rm dis}/{\rm yr})}=10.2$--$10.8$.

\begin{table*}\centering
\caption[]{\label{tab:parameters}\sf
     Adopted parameters. For any SFR and $\Sigma_{\rm SFR}$, Eqs.~\ref{eq:cfe} and~\ref{eq:mstar} give the values of $\Gamma$ and $M_{\rm c}$. Only their resulting dependences on $M_{\star,{\rm gal}}$ are listed here.}
\begin{tabular}{l c c c c c c}
\hline
{\sf SF mode} & $f_{\star,{\rm old}}$ & ${\rm SFR}~(\msun~{\rm yr}^{-1})$ & $\Sigma_{\rm SFR}~(\msun~{\rm yr}^{-1}~{\rm kpc}^{-2})$ & $\Gamma$ & $M_{\rm c}$ & $C_{\rm dis}~({\rm yr})$
\\\hline
{\sf Extreme} & $1$ & - & - & $1$ & $\infty$ & $\infty$  \\
{\sf Rapid burst} & $1$ & $32.9(M_{\star,{\rm gal}}/10^9~\msun)^{0.96}$ & $4.53(M_{\star,{\rm gal}}/10^9~\msun)^{0.60}$ & $\propto M_{\star,{\rm gal}}^{0.14}$ & $\propto M_{\star,{\rm gal}}^{0.48}$ & $10^{10.8}$  \\
{\sf Mild burst} & $0.8$ & $0.28(M_{\star,{\rm gal}}/10^9~\msun)^{0.80}$ & $0.033(M_{\star,{\rm gal}}/10^9~\msun)^{0.41}$ & $\propto M_{\star,{\rm gal}}^{0.10}$ & $\propto M_{\star,{\rm gal}}^{0.4}$ & $10^{10.5}$ \\
{\sf Quiescent} & $0.6$ & $M_{\star,{\rm gal}}/\tau$ & $0.021(M_{\star,{\rm gal}}/10^9~\msun)^{0.75}$ & $\propto M_{\star,{\rm gal}}^{0.18}$ & $\propto M_{\star,{\rm gal}}^{0.5}$ & $10^{10.2}$   \\
\hline
\end{tabular} 
\end{table*}
By combining the above set of equations with Eq.~\ref{eq:pmmax}, we can relate the galaxy stellar mass to the most massive GC that is expected if the GCs were formed with the same ICMF as YMCs in the nearby universe. For this, we adopt certain relations or values for SFR$(M_{\star,{\rm gal}})$, $\Sigma_{\rm SFR}(M_{\star,{\rm gal}})$, $f_{\star,{\rm old}}$ and $C_{\rm dis}$. Due to the dependence on the SFR and $\Sigma_{\rm SFR}$, we need to define two limiting modes of GC formation that span the range of plausible star formation histories (SFHs) of the galaxies. We use the semi-analytic galaxy models of \citet{guo11}, based on the Millennium-II simulation\footnote{See http://www.mpa-garching.mpg.de/millennium} \citep{boylankolchin09}. This then spans the range  in the $M_{\rm max}$--$M_{\star,{\rm gal}}$ plane that would be expected if GCs were formed according to the ICMF of YMCs. The two limiting modes are:
\begin{itemize}
\item[(1)] The `quiescent mode', which assumes that the SFR has been constant throughout the history of the galaxy. In this mode, $f_{\star,{\rm old}}$ is set by the definition that any star or cluster formation corresponding to ages $\tau>5$~Gyr is considered to be coeval with GC formation. { While this age limit is arbitrary to some degree, it ensures that we are considering the galaxies at times when GCs may have formed.} We have fitted $\Sigma_{\rm SFR}(M_{\star,{\rm gal}})$ by selecting galaxies from \citet{guo11} at the moments when they are forming stars at a rate lower than their mean SFRs. { This selection guarantees that any episodes of enhanced star formation are excluded (as, by definition, the SFR during these episodes will be greater than the average SFR over the history of the galaxy). For a given galaxy, $\Sigma_{\rm SFR}$ is obtained by dividing its SFR by the area within its half-mass radius at each quiescent time-step, and then taking the average of these values. This approach accounts for the relation between the SFR and $\Sigma_{\rm SFR}$ as the structure of the galaxy changes. $\Sigma_{\rm SFR}(M_{\star,{\rm gal}})$ is then obtained by fitting a power law relation to the galaxy sample.} We also adjust $C_{\rm dis}$ to the low end of its range of possible values. {\it This mode leads to the lowest maximum GC masses.}
\item[(2)] The `rapid burst mode', which assumes that all stars in the galaxy were formed coevally with the GCs in a starburst. { This implies $M_{\star,{\rm gal}}\equiv M_{\rm burst}$, with $M_{\rm burst}$ the total mass formed in the burst. We have fitted ${\rm SFR}(M_{\rm burst})$} by { again selecting galaxies} at ages $\tau>5$~Gyr from \citet{guo11}, { this time picking starbursts (i.e. the times at which the SFR of each galaxy peaks)} with durations shorter than the time resolution (300~Myr). We then assume burst durations of one dynamical time { to determine the actual SFR}, i.e. $t_{\rm burst}=3R/V$ with $R$ the disc scale length and $V$ the rotational velocity. Again, the area covered by the half-mass radius is used to determine $\Sigma_{\rm SFR}$. We adjust $C_{\rm dis}$ to the high end of its range of possible values. {\it This mode leads to the highest maximum GC masses.}
\end{itemize}
The resulting parameters and proportionalities for $\Gamma(M_{\star,{\rm gal}})$ and $M_{\rm c}(M_{\star,{\rm gal}})$ are shown for all scenarios in Table~\ref{tab:parameters}. In addition to the above two modes of GC formation, we include an intermediate, `mild burst' scenario for which we have fitted ${\rm SFR}(M_{\star,{\rm gal}})$ and $\Sigma_{\rm SFR}(M_{\star,{\rm gal}})$ to galaxies from \citet{guo11} at the moments when they are forming stars at a rate higher than their mean SFRs, { analogously to the quiescent mode above. This selection ensures that only episodes of enhanced star formation are included.}

\begin{figure}
\center\resizebox{8cm}{!}{\includegraphics{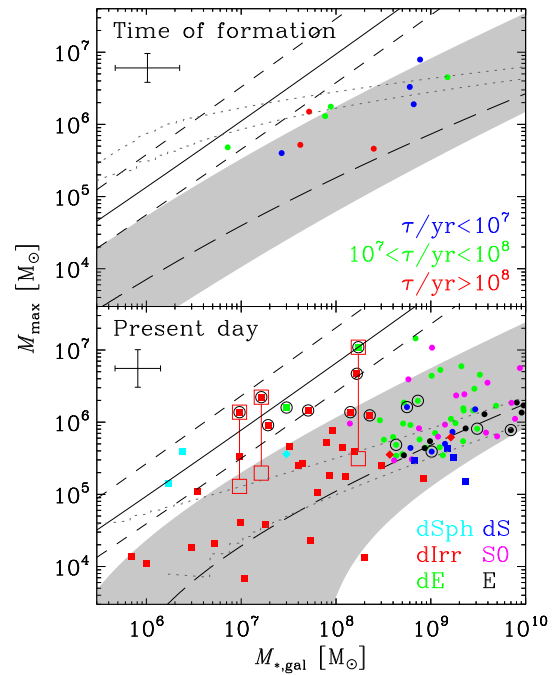}}\\
\caption[]{\label{fig:mgcmax}\sf
      Mass of the most massive star cluster versus the stellar mass of its host galaxy. Solid and dashed lines indicate the upper limit and its standard deviation from Fig.~\ref{fig:mgcmax0}. The shaded area represents the parameter space encompassed by the `quiescent mode' and `rapid burst mode' from Table~\ref{tab:parameters}, which represent the physically plausible extremes. The region includes one standard deviation up and down. The long-dashed line indicates the `mild burst mode'. The dotted lines mark the equivalent of the shaded area for a log-normal ICMF (see Sect.~\ref{sec:method}). Top: initial relation, excluding mass loss due to stellar evolution and tidal disruption. Symbols indicate YMCs in the nearby universe, coloured by their ages $\tau$. Their location on the x-axis is determined by the total stellar mass formed coevally with the YMC (see text). Bottom: relation at the present day, after the application of mass loss due to stellar evolution and tidal disruption. Symbols represent the sample of GCs and their host galaxies as in Fig.~\ref{fig:mgcmax0}. The typical error margins on the data are shown in the top left corners of each panel. GCs that are more than 2$\sigma$ separated from the `allowed' area are marked with red boxes, { which are connected by red lines to the second most massive GCs in these galaxies.}
                 }
\end{figure}
\section{The formation of globular clusters} \label{sec:obs}
The region in the $M_{\star,{\rm gal}}$--$M_{\rm max}$ plane that can be spanned by random sampling from a Schechter-like CIMF for the physically plausible range of dwarf galaxy histories is shown in Fig.~\ref{fig:mgcmax}. The top panel shows the relation at the time of GC formation and includes the most massive YMCs of nearby galaxies\footnote{In order of increasing coeval stellar mass, the galaxies denoted by the symbols are NGC1705 (ref. 1,2), Mrk930 (ref. 3), M83 (ref. 4,5), NGC4214 (ref. 5,6), NGC1569 (ref. 1,7), NGC6946 (ref. 4,5), NGC4449 (ref. 5,6), NGC4038/39 (ref. 8,9,10), NGC3256 (ref. 11,12), Haro11 (ref. 13,14) and M82 (ref. 15,16). The sources that were used to compile the data are: (1) \citet{smith01}, (2) \citet{annibali03}, (3) \citet{adamo11}, (4) \citet{larsen00b}, (5) \citet{larsen04c}, (6) \citet{mcquinn10}, (7) \citet{grocholski08}, (8) \citet{zhang01}, (9) \citet{mengel02}, (10) \citet{karl10}, (11) \citet{zepf99}, (12) \citet{trancho07}, (13) \citet{hayes07}, (14) \citet{adamo10}, (15) \citet{foerster03}, (16) \citet{konstantopoulos09}.} for reference. { For these data points, the position on the x-axis does not reflect} the total stellar mass { of the galaxies}, but merely the mass that was formed coevally with the most massive cluster -- either given by the integral of the SFR over the starburst or, if the SFR has been constant or the SFH is unknown, the product of the current SFR with twice the age of the YMC. The YMCs are scattered along the top edge of the `allowed' region, because most of them reside in starburst galaxies and we have implicitly assumed $f_{\star,{\rm old}}=1$ by using the coeval stellar mass. Acknowledging this, we see that the YMCs fall in the anticipated range for the quiescent and rapid burst modes, which illustrates that our approach is indeed consistent with their formation. Due to their Schechter-type ICMF, the YMCs are not consistent with the relation for a log-normal ICMF.

{ The bottom panel of Fig.~\ref{fig:mgcmax} shows the observed and predicted present-day masses of the most massive GCs in the dwarf galaxies of our sample as a function of the total host galaxy stellar mass.} In this panel, mass loss due to stellar evolution and tidal disruption has been included, which skews the region that is spanned by the quiescent and rapid burst modes and moves it to lower masses. There is good agreement with the observed GCs, suggesting that the formation of GCs is indeed consistent with the same ICMF as YMCs. It also indicates that the CFE and $M_{\rm c}$ followed the same relations for GCs as they do for YMCs. { If the ICMF were not (exponentially) truncated, most of the Virgo Cluster galaxies would fall well below the predicted region.}

{ At least some discrepant GCs are expected statistically, and indeed a handful of GCs have masses that are 1$\sigma$ to 2$\sigma$ outliers from the predicted range. For the two dSph galaxies at $M_{\star,{\rm gal}}\sim 2\times10^6~\msun$, this could have a physical explanation, such as a very efficient starburst that formed the entire galaxy at a CFE of $\Gamma=1$. Looking at statistically more substantial extremes, we see that} three out of the 101 GCs have masses that are more than $2\sigma$ separated from the `allowed' region in the parameter space (marked by red boxes in Fig.~\ref{fig:mgcmax}), whereas a { purely} log-normal ICMF is more than $2\sigma$-inconsistent with the masses of 16 GCs, and generally cannot reproduce the observed spread in the $M_{\rm max}$--$M_{\star,{\rm gal}}$ relation. When leaving out nuclear GCs (nGCs, see below), a total of eight $2\sigma$-inconsistent GCs are left, ruling out a log-normal ICMF for metal-poor GCs in dwarf galaxies at a 99.6\% confidence level. By contrast, there are no non-nuclear outliers at all for the Schechter-type ICMF. { We have tested the agreement for log-normal ICMFs with different parameters and find that a log-normal} can only yield sufficiently high GC masses for dispersions $\sigma>1.4$ \citep[which is inconsistent with][]{vesperini03}, but this would still not reproduce the large spread in $M_{\rm max}$. { The possibility of a higher peak mass is ruled out by current GC systems, because dynamical evolution is not capable of then reducing the peak mass to its current value\footnote{{ Unless massive clusters are more rapidly disrupted than low-mass ones, contrary to what is shown by theory and observations \citep[e.g.][]{lamers10,bastian11b}.}} \citep{jordan07,gieles09}.}

All three GCs that are more than 2$\sigma$-inconsistent with sampling from a Schechter-type ICMF are nGCs (black circles in Fig.~\ref{fig:mgcmax}). { In these galaxies, the second most massive GCs are consistent with the predicted mass range as well as with the masses of the GCs in the other galaxies. Their masses require them to have formed in the rapid burst mode, especially since they are only the second most massive GCs. The abundance of nGCs among the outliers} is not necessarily surprising, since nGCs are thought to be the products of a combination of merged star clusters that migrated to the galaxy centre due to dynamical friction \citep{miocchi06} and additional gas accretion \citep{hartmann11}. This allows their mass to increase beyond the limit imposed by the finite duration of the star formation process. The form of the ICMF is the outcome of the hierarchical merging of star-forming aggregates \citep{elmegreen96}, and is therefore normalised to the total mass forming in stellar clusters. By contrast, the nGCs are consistent with the `extreme mode' of random sampling from Fig.~\ref{fig:mgcmax0}, in which the stellar mass of the entire galaxy is converted into an ICMF. They could thus be regarded as the high-mass end of the mass spectrum of all (stellar) structure in the galaxy, driven by hierarchical merging over a Hubble time rather than on a star formation time-scale.

The fraction of galaxies with a $2\sigma$-outlier as their most massive GC is $\sim 10$\% between $M_{\star,{\rm gal}}=10^7$--$10^9~\msun$, and decreases for more massive galaxies. Since nGCs are likely formed through dynamical friction \citep{bellazzini08}, this is probably caused by the increase of the dynamical friction time-scale with host galaxy mass, implying that the orbital decay of GCs and their potential merging becomes less efficient in more massive galaxies. Indeed, \citet{bekki10} identify a disc mass of about $10^9~\msun$ as the upper limit for efficient GC migration due to dynamical friction and corresponding nGC growth, in good agreement with the data.

The dwarf galaxy stellar mass range $M_{\star,{\rm gal}}=10^7$--$10^9~\msun$ is of particular interest because the stellar halo of the Milky Way is thought to have formed by the accretion of galaxies with such masses \citep{cooper10}. If we account for the disruption history of GCs, this can be used to give a first-order estimate for the fraction of the current metal-poor Galactic GCs have formed in a similar way to YMCs, and which fraction formed as an nGC before they were accreted. The masses of outlier nGCs in Fig.~\ref{fig:mgcmax} are about an order of magnitude higher than other GCs in current dwarf galaxies, but if they were accreted earlier they may not have had the time to grow as much. Taking an intermediate mass difference of a factor of five, Eq.~\ref{eq:lamers} shows that the disruption rate of nGCs is about three times lower than for GCs. We can then use the difference in the specific frequency of GCs between dwarf galaxies and Milky Way-mass galaxies in the Virgo Cluster \citep[number of metal-poor GCs per unit galaxy stellar mass $T_{\rm blue}$, with $T^{\rm spiral}_{\rm blue}/T^{\rm dwarf}_{\rm blue}\sim 0.2$, see][]{peng08} to get a rough idea of the total amount of disruption during the assembly of the Galactic GC system from accreted dwarf galaxies. Combining this with the initial nGC fraction of 10\%, we find that the current fraction of { metal-poor} Galactic GCs that was formed differently than YMCs could well be as high as 35\%. { This fraction is an upper limit because it assumes that all current metal-poor Galactic GCs were once the most massive cluster of a dwarf galaxy, which need not be the case. Since metal-poor GCs constitute about half of the Galactic GC population, the Galaxy-wide upper limit to the number of GCs with a nuclear origin is 15--20\%.}

{The fraction of GCs with a nuclear origin is interesting in view of the several Galactic GCs with chemical signatures that are strongly inconsistent with single stellar populations \citep[e.g.][]{gratton04}. If these chemical anomalies can be traced to former nGCs \citep{carretta10}, the observed distinct stellar populations would require them to be the product of one or two GC-GC mergers in the centre of their host galaxies instead of a continuous accretion. This is not unlikely, because dynamical friction is most efficient for the most massive clusters. {According to current data, the fraction of Galactic GCs with the light element abundance anomalies that indicate multiple stellar populations is much larger than 35\% \citep{caloi11}, casting doubt on the feasibility of explaining them by nGCs. Instead, it suggests that light element abundance anomalies should also be present in the most massive YMCs in the nearby universe -- depending on how long it takes for a second generation of stars to form. However, the fraction of GCs with Fe abundance variations is smaller \citep[e.g.][]{gratton04}, and could possibly be consistent with the estimated fraction of nGCs.}

We have shown that the majority (65--90\%) of the metal-poor GCs in dwarf galaxies are consistent with the Schechter-type ICMF of YMCs, which is required for their formation mechanisms to have been the same. A log-normal ICMF is ruled out with 99.6\% significance. However, in about 10\% of the dwarf galaxies between $M_{\star,{\rm gal}}=10^7$--$10^9~\msun$ the masses of the most massive GCs are at least $2\sigma$ too high to be consistent with YMCs. These outliers are generally nGCs, of which the masses have increased due to the merging with other clusters and further gas accretion, allowing them to exceed the mass-scales set by the star formation process (and thus the ICMF). Their elevated masses imply a higher chance to survive any tidal disruption during and since their accretion onto the Milky Way than `normal' GCs, suggesting that the current fraction of metal-poor GCs with a nuclear origin will have risen above 10\%, potentially up to 35\%.

\section*{Acknowledgments}
We are grateful to the anonymous referee for a helpful and constructive report. We thank Nate Bastian, Eli Bressert, Mark Gieles, Nora L\"{u}tzgendorf, Simon White and Tim de Zeeuw for valuable comments on the paper, and Iskren Georgiev and Markus Kissler-Patig for helpful discussions. The Millennium-II Simulation databases used in this paper were constructed as part of the activities of the German Astrophysical Virtual Observatory.

\bibliographystyle{mn2e2}
\bibliography{mybib}

\bsp

\label{lastpage}

\end{document}